\documentclass[aps,twocolumn,pre,showpacs,longbibliography]{revtex4-1}
\usepackage{graphicx}
\usepackage{amsmath}
\usepackage{amsfonts}
\usepackage[table]{xcolor}
\usepackage[caption=false]{subfig}
\usepackage{array}
\usepackage{floatrow}
\usepackage{pbox}
\usepackage{physics}
\usepackage[colorlinks=true,linkcolor=blue,citecolor=blue,filecolor=cyan,urlcolor=blue,breaklinks=true]{hyperref}

\begin{document}
	\title{Optimality of non-conservative driving for finite-time processes with discrete states}
	
	\author{Benedikt Remlein and Udo Seifert}
	\address{ II. Institut f\"ur Theoretische Physik, Universit\"at Stuttgart,
		70550 Stuttgart, Germany}
	\date{\today}
	
	\parskip 1mm

	\begin{abstract}
		An optimal finite-time process drives a given initial distribution to a given final one in a given time at the lowest cost as quantified by total entropy production. We prove that for system with discrete states this optimal process involves non-conservative driving, i.e., a genuine driving affinity, in contrast to the case of system with continuous states. In a multicyclic network, the optimal driving affinity is bounded by the number of states within each cycle. If the driving affects forward and backwards rates non-symmetrically, the bound additionally depends on a structural parameter characterizing this asymmetry.
	\end{abstract}
	
	%\pacs{05.70.Ln, 05.40.-a}
	% Explanation of PACS numbers:
	% 05.70.Ln: Nonequilibrium and irreversible thermodynamics
	% 05.40.-a: Fluctuation phenomena, random processes, noise, and Brownian motion

	\maketitle

	In thermodynamics, a finite-time process
	transforms a given initial state into a given
	final one in a given finite time. This process is optimal if it comes at the lowest cost, i.e., at the lowest entropy
	production. The condition of a finite time
	is crucial, since quasi-static processes,
	which require infinitely slow driving, do not
	generate entropy at all. For macroscopic systems, such processes have been studied under the label of finite-time thermodynamics \cite{andr11}.
	For small systems
	in contact with a thermal environment and thus following a stochastic dynamics, optimal finite-time processes were shown 
	to have an	inevitable thermodynamic cost that scales 
	asymptotically like the inverse of the allocated time \cite{seki97a, schm07}. This scaling was later shown to be 
	the exact minimal entropy production for any
	finite time for an underlying Langevin dynamics \cite{schm08, aure12}. 
	For system with discrete state space undergoing a master equation dynamics, this scaling holds asymptotically as several case studies have shown \cite{espo10a,dian13,zulk11}.
	In the linear response regime, 
	an appealing systematic theory for the optimal driving involves
	geometric concepts like the thermodynamic length  \cite{croo07,siva12, zulk12, mura13, zulk15, siva16}. For an effective two-state system, a prominent experimental application of optimal
	protocols is the minimal cost of erasing a bit in a finite-time extension of the
	Landauer bound \cite{beru12,jun14, proe20}.
		
	A fundamental distinction for any non-equilibrium process is whether or not
	the driving is conservative, i.e., 
	whether or not it
	arises from a time-dependent
	potential.
	The former case applies \textsl {inter alia} to single molecules
	manipulated  with optical tweezers \cite{wood14,camu16}, to
	colloidal particles in time-dependent 
	harmonic or anharmonic traps \cite{cili17}  and to stochastic pumps for which the energy of each state 
	(and potentially the barriers in between)
	are driven \cite{sini07,raha08,cher08}. Paradigms for non-conservative driving are colloidal particles driven along static periodic potentials and colloidal particles in shear flow. Biophysical and biochemical processes that are driven by unbalanced chemical reactions like the hydrolysis of nucleic acids fall in this class as well \cite{fang19,mugn20}.
	
	Emphasizing this distinction leads to the
	question whether conservative or non-conservative driving leads to a lower
	cost for a given initial and final state. In a more technical formulation, the
	question is whether a time-dependent dynamics whose instantaneous stationary state is Boltzmann-Gibbs-like achieves 
	already minimal entropy production
	or whether an additional non-conservative contribution, which at fixed control parameter would lead to a genuine non-equilibrium steady-state, can further decrease the cost. For systems with a continuous state space, i.e., for Langevin dynamics, 
	it is known that the optimal protocol
	involves only conservative forces  \cite{aure11,aure12,gawe13,dech19}.
	Coming back to the example of a colloidal particle on a ring with periodic boundary conditions this result implies that there is nothing gained by allowing a non-conservative force
	to act on top of a time-dependent potential.
	
	In this Letter, we address this question for systems with discrete states, i.e., for
	a master equation dynamics with time-dependent rates. 
	We can build on the work of Muratore-Ginanneschi et al. \cite{mura12} who formulated this optimization problem in terms of 
	control theory without addressing the specific question 
	we are interested in. Since they show that the
	optimization reduces to the Langevin problem in the continuum limit, one might even expect that
	the optimal protocol for a discrete state space
	can be achieved with conservative driving as well. 
	In contrast to the
	continuous case, however, we will prove that the
	optimal driving is in fact non-conservative. Furthermore, we will show that for a broad class of systems 
	all cycle affinities, defined precisely below as a quantitative measure of
	the "non-conservativeness" of the dynamics, remain bounded as a
	function of the number of states in a cycle during the
	whole process independent of its duration.

	We consider a discrete set of states $\{i\}$ of total number $N$. A transition between two states $(i,j)$ occurs at a rate $k_{ij}(t)$, which is in general time-dependent. The probability $p_i(t)$ to find the system at time $t$ in state $i$ evolves according to the master-equation
	\begin{equation}
		\partial_t p_i(t) = \sum_{j\neq i} [k_{ji}(t)p_j(t) - k_{ij}(t)p_i(t)] \equiv -\sum_{j\neq i} j_{ij}(t)
		\label{eq:masterK}
	\end{equation}
	with the net probability current $j_{ij}(t)$ through link $(i,j)$. We parameterize the transition rates as \cite{mura12}
	\begin{equation}
		k_{ij}(t) = \kappa_{ij}e^{A_{ij}(t)/2}
		\label{eq:Rates}
	\end{equation}
	with constant, symmetric part $\kappa_{ij}= \kappa_{ji}$, which sets the characteristic time-scale for the transition from state $i$ to $j$, and time-dependent, antisymmetric $A_{ij}(t) = - A_{ji}(t)$. 
	Throughout the paper, we measure energies in units of a thermal energy and entropy in units of Boltzmann's constant.
	
	Conservative driving implies that the ratio of forward and backward rates is given by the difference of time-dependent
	free energies $F_i(t)$ leading to
	\begin{equation}
		A_{ij}(t)=F_i(t)-F_j(t).
	\end{equation}
	In contrast, for non-conservative driving $A_{ij}(t)$ cannot
	be written as a difference of state functions.
	
	It will be convenient to transform the state densities as $\phi_i(t) \equiv \sqrt{p_i(t)}$ and to introduce the non-equilibrium driving function \cite{mura12}
	\begin{equation}\begin{split}
			\varphi_{ij}(t) &\equiv A_{ij}(t) + 2[ \ln \phi_i(t) - \ln \phi_j(t)]~,\end{split}\label{eq:Driving}\end{equation}
	which becomes for conservative driving
	\begin{equation}\begin{split}
			\varphi_{ij}(t) &= B_i(t)-B_j(t)~~~\textrm{with}
			~~~ B_i(t)\equiv  F_i(t)+2 \ln \phi_i(t) .
		\end{split}
		\label{eq:Driving2}\end{equation}	
	In this representation, the current along link $(i,j)$ transforms to 
	\begin{equation}j_{ij}(t) = 2\kappa_{ij}\phi_i(t) \phi_j(t) \sinh \frac{\varphi_{ij}(t)}{2}
		\label{eq:current}
	\end{equation}
	and the master equation (\ref{eq:masterK}) becomes
	\begin{equation}
		\partial_t \phi_i(t) = - \sum_{j\neq i} \kappa_{ij} \phi_j(t) \sinh \frac{\varphi_{ij}(t)}{2}.
		\label{eq:master}\end{equation}
	The process that transforms the given initial density $\{\phi_i(0)\}$ to a given final one $\{\phi_i(T)\}$
	in time $T$ is optimal if it generates the least overall entropy production
	\begin{equation}
		\Delta S_{\rm{tot}} \equiv \int_0^T \text{d}t~\sigma(t)~
		\label{eq:totalS}
	\end{equation}
	with the entropy production rate \cite{seif12}
	\begin{equation}
		\begin{split}
			\sigma(t) & =\sum_{i\neq j} p_i(t) k_{ij}(t) \ln \frac{p_i(t) k_{ij}(t)}{p_j(t) k_{ji}(t)}\\
			& = 2\sum_{i<j}\kappa_{ij} \phi_i(t) \phi_j(t) \varphi_{ij}(t)\sinh\frac{\varphi_{ij}(t)}{2}~.
		\end{split}
		\label{eq:EPR}
	\end{equation}
	
	We first show that conservative driving does not lead to
	minimal entropy production. Assume that 
	within this parameter space (\ref{eq:Driving2}) we have found
	the optimal protocol $F^*_i(t)$ 
	leading to $p_i^*(t)$ with currents $j^*_{ij}(t)$.
	For a unicyclic system,  it is then
	clear that adding a time-dependent $\Delta(t)$ to
	the clockwise current will still satisfy the master equation (\ref{eq:masterK})
	and the boundary conditions of a fixed initial and final density. Under the transformation 
	\begin{equation}
		j_{ij}(t) \equiv j^*_{ij}(t) + \epsilon_{ij}\Delta(t)\label{eq:trafo}\end{equation}
	with $\epsilon_{ij} = 1 = -\epsilon_{ji}$ for $i<j$ the  entropy production rates $\sigma^*(t)$, respectively $\sigma(t)$, become by a Taylor-expansion
	\begin{equation}\begin{split}
			\sigma(t)- \sigma^*(t)  &= \\&\quad\Delta(t) ~ 2\sum_{i<j} \tanh\frac{B^*_i(t)-B^*_j(t)}{2}+\mathcal O(\Delta(t) ^2)~.
			\label{eq:bSigmaFirst}
	\end{split}\end{equation}
	Since in general, the linear term will not vanish,
	we get that the total entropy production found
	within conservative driving can be further decreased by adding a non-conservative term accounting for such a $\Delta (t)$.
	Specifically, we can choose $\Delta(t) = const. \neq 0$ such that \begin{equation}
	2	\Delta ~ \int_0^T \text{d}t  \sum_{i<j} \tanh\frac{B^*_i(t)-B^*_j(t)}{2} < 0~.
	\end{equation} 
	This constitutes our first main result: In contrast to the continuous case, optimal protocols for Markov jump processes involve non-conservative driving, i.e., a genuine cycle affinity
	\begin{equation}
		\mathcal A_c(t) \equiv \sum_{(i,j)\in 
			\mathcal{C}} A_{ij}(t) = \sum_{(i,j)\in \mathcal{C}} \varphi_{ij}(t)
	\end{equation}
	for each cycle $\mathcal{C}$ in the network. The above proof can indeed be extended trivially to multicyclic networks since a corresponding $\Delta(t)$ can be added to an
	arbitrary cycle in which case the summation in 
	(\ref{eq:bSigmaFirst}) is only over the directed links of this cycle. 
	
	We next show that all cycle affinities $\mathcal A_c(t)$
	are bounded by the number of states in each cycle
	for all times. To do so, we have to derive the
	Euler-Lagrange equations for the variational problem
	posed by minimizing the entropy production (\ref{eq:totalS}) under
	the constraints (\ref{eq:master}) which we add with Langrangean multipliers
	$\{\eta_i(t)\}$ that ensure that the densities $\{\phi_i(t)\}$ satisfy the master equation.
	Thus, we minimize 
	\begin{equation}
		\Delta S  \equiv \int_0^T \text{d}t~L[\{\phi(t),\varphi(t)\}]~
		\label{eq:functional}
	\end{equation}
	 for given $\{\phi_i(0)\}$ and $\{\phi_i(T)\}$
	 with Lagrange function
	\begin{equation}
		L(t) \equiv \sigma(t) + \sum_i \eta_i(t) [ \partial_t \phi_i(t) + \sum_{j\neq i} \kappa_{ij}\phi_j(t)\sinh\frac{\varphi_{ij}(t)}{2}]~.
	\end{equation}
	From ${\delta L}/{\delta \phi_i(t)}=0$, we get the equations of motion for the Lagrange multiplier
	\begin{equation}
		\partial_t \eta_i(t) = \sum_{j\neq i}\kappa_{ij} \sinh\frac{\varphi_{ij}(t)}{2} [2 \phi_j(t) \varphi_{ij}(t) - \eta_j(t)]~.
		\label{eq:eta}
	\end{equation}
	Variation with respect to the protocol $\varphi_{ij}(t)$ leads to
	\begin{equation}
		\eta_i(t) \phi_j(t) - \eta_j(t) \phi_i(t) = -4 \phi_i(t) \phi_j(t) [\tanh \frac{\varphi_{ij}(t)}{2}+ \frac{\varphi_{ij}(t)}{2}]~.
		\label{eq:varphi}\end{equation}
	
	By summing (\ref{eq:varphi}) over an arbitrary cycle with $N_c$ states we get
	\begin{equation}\begin{split}
			0 &=\sum_{i=1}^{N_c} [ \phi_i(t)/\eta_i(t) - \phi_{i+1}(t)/\eta_{i+1}(t)]\\
			&= - 4\sum_{i=1}^{N_c} [ \tanh\frac{\varphi_{i,i+1}(t)}{2}+ \frac{\varphi_{i,i+1}(t)}{2}]
		\end{split}
		\label{eq:tanhSum}\end{equation}
	where we relabeled the neighboring links $(i,j) \in \mathcal C$ as $(i,i+1)$. We now use this relation to find for the affinity
	\begin{equation}\begin{split}
			\mathcal A_c(t) &= \sum_{i=1}^{N_c}\varphi_{i,i+1}(t) = -2 \sum_{i=1}^{N_c} \tanh\frac{\varphi_{i,i+1}(t)}{2}~
		\end{split}\label{eq:affinity}\end{equation}
	and finally use $|\tanh(x)| \leq 1$ to obtain
	\begin{equation}
		|\mathcal A_c(t)|  \leq 2  N_c~.
		\label{eq:bound}
	\end{equation}
	Thus for each cycle, the time-dependent affinity is bounded by the number of states within that cycle.
	
	In fact, we can sharpen this bound further to 
	\begin{equation}
		|\mathcal A_c(t)| \leq 2 (N_c-2).
		\label{eq:sharpBound}
	\end{equation}
	which is our second main result. While the formal derivation of this improved bound as shown in \cite{SM} is somewhat technical, its origin can be understood by the following consideration. 
	Eqs. (\ref{eq:affinity},\ref{eq:bound}) require the affinity and hence the sum of the driving functions to be finite, thus, not all $\{\varphi_{ij}(t)\}$ are allowed to tend to, e.g., positive infinity at the same time which was the rational behind the weaker bound, Eq. (\ref{eq:bound}). At least one driving function has to compensate this putative divergence by approaching negative infinity. The asymptotic behavior of the affinity is determined by Eq. (\ref{eq:tanhSum}), thus, for all $\varphi_{ij}(t) \to \pm \infty$ except one that tends to $\varphi_{kl}(t) \to \mp \infty$, the affinity approaches $\mathcal A_c(t) \to \mp 2(N_c -2)$.

	We now turn to numerics in order to explore how significant the improvement through non-conservative driving
	is and to check how strong the improved bound (\ref{eq:sharpBound}) is. For a three state system, i.e., $N_c=3$, we sample arbitrary initial and final distributions and calculate for each pair of
	them the optimal protocol first for conservative and then
	for non-conservative driving. We fix a basic time-scale by setting all symmetric prefactors $\kappa_{ij} = \kappa_{ji} = 1$.
	We find that the non-conservative driving leads to an only minute improvement. For a process transforming the state of the system
	within a time that is comparable to the intrinsic timescale, i.e., for $T = 1$, the advantage of non-conservative
	driving is on average only of the order of $10^{-5}$ with a maximal improvement of order $10^{-4}$. Even for processes
	that are ten times faster, i.e., $T = 0.1$, on average this advantage raises only to $10^{-3}$, respectively $10^{-2}$ for the maximum value.
	
%
%Mean dS1 =  3.137706762443116e-05
%Median dS1 =  1.0034888234069757e-05
%sigma dS1 =  5.061670035571342e-05
%Max dS1 =  0.00035449286487608365
%Mean dS01 =  0.002676717244154088
%Median dS01 =  0.001853527360126268
%sigma dS01 =  0.0025257646933047324
%Max dS01 =  0.010976588988499117	
	
	\begin{figure}
		\centering
		\includegraphics[width=0.75\textwidth]{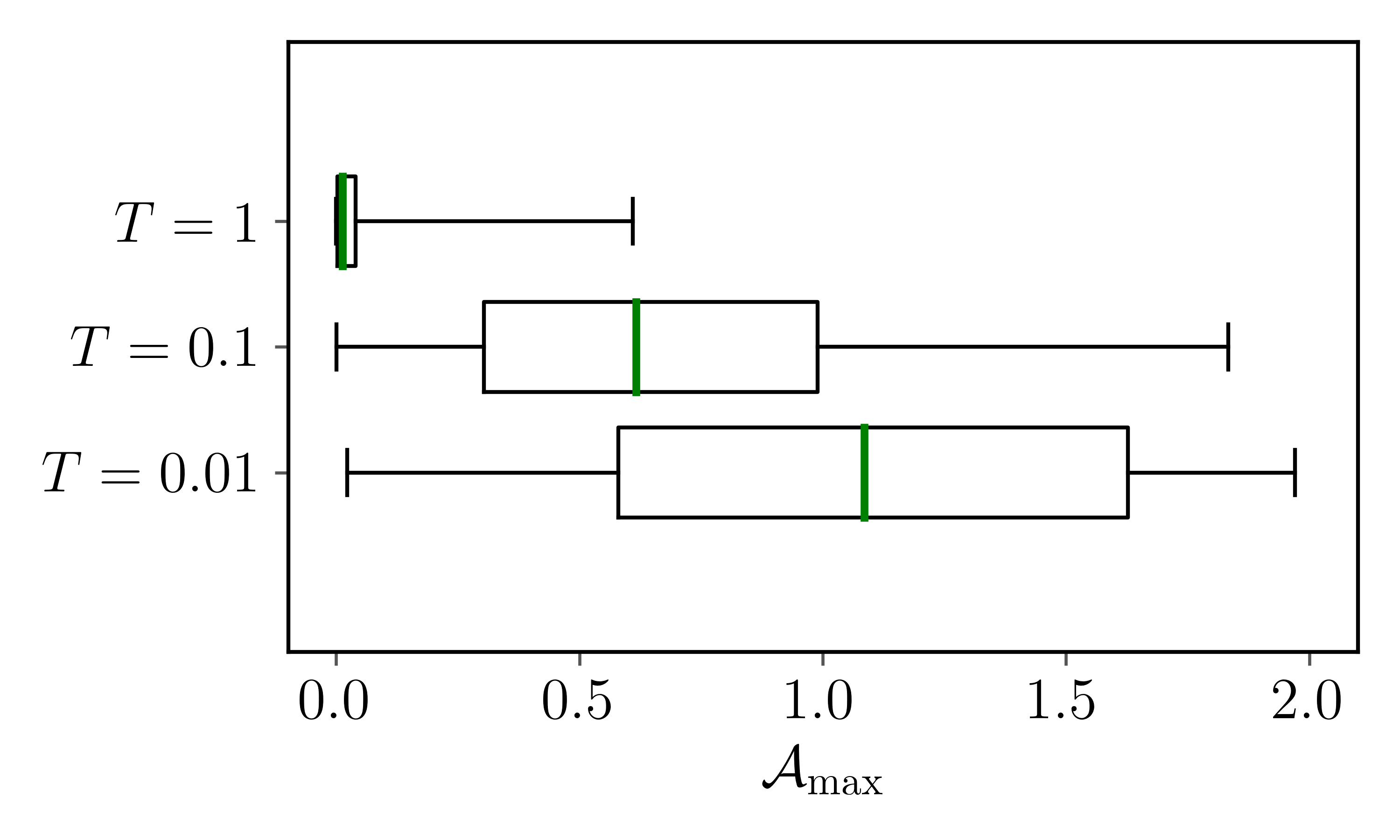}
		\caption{Influence of the allocated time $T$ on the maximal affinity $\mathcal A_{\rm{max}}$ for arbitrary sampled initial and final distributions. The edges of the boxes represent the first $Q_1$ (left) and third quartile $Q_3$ (right site). The whisker on the left represents the minimum value and on the right the maximum of the data. The thick green line displays the median of the data.}
		\label{fig:Amax}
	\end{figure}
	
\begin{figure}[t]
	\subfloat[$T =1$ \label{sfig:t1All}]{%
		\includegraphics[width=0.35\textwidth]{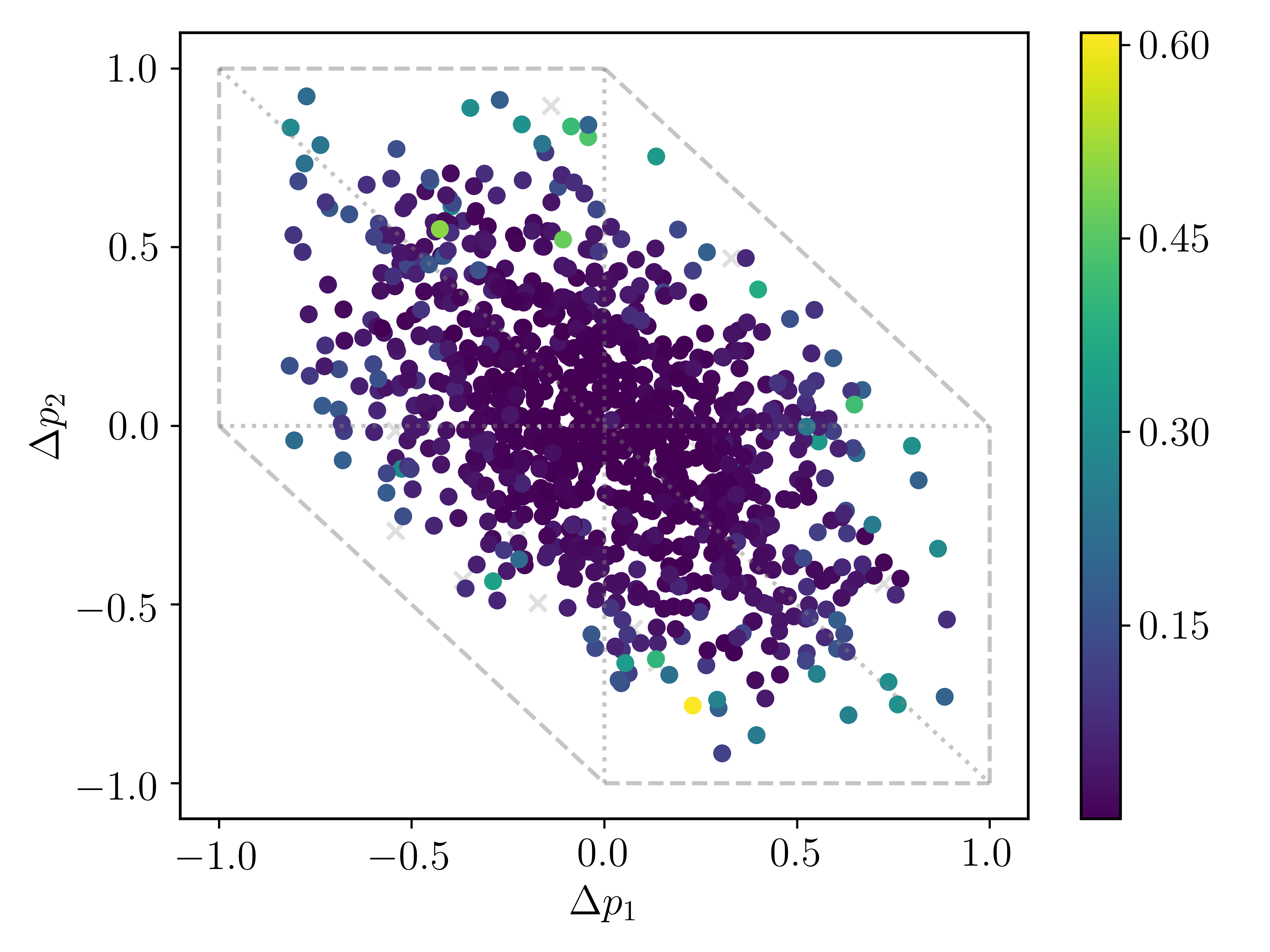}
	}%\hfill
	\subfloat[$T =0.1$ \label{sfig:t01All}]{%
		\includegraphics[width=0.35\textwidth]{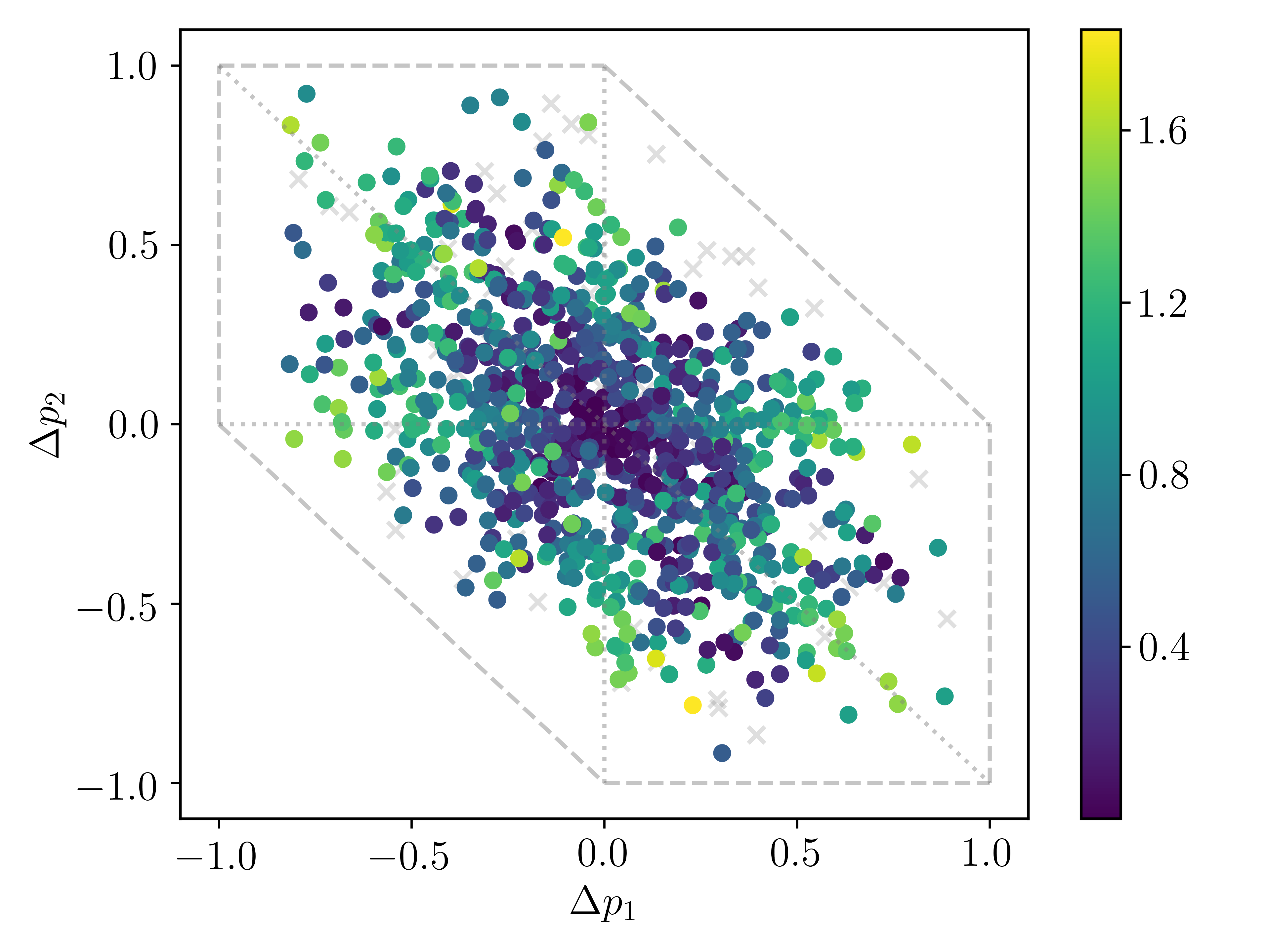}
	}%\hfill
	\subfloat[$T =0.01$ \label{sfig:t001All}]{%
		\includegraphics[width=0.35\textwidth]{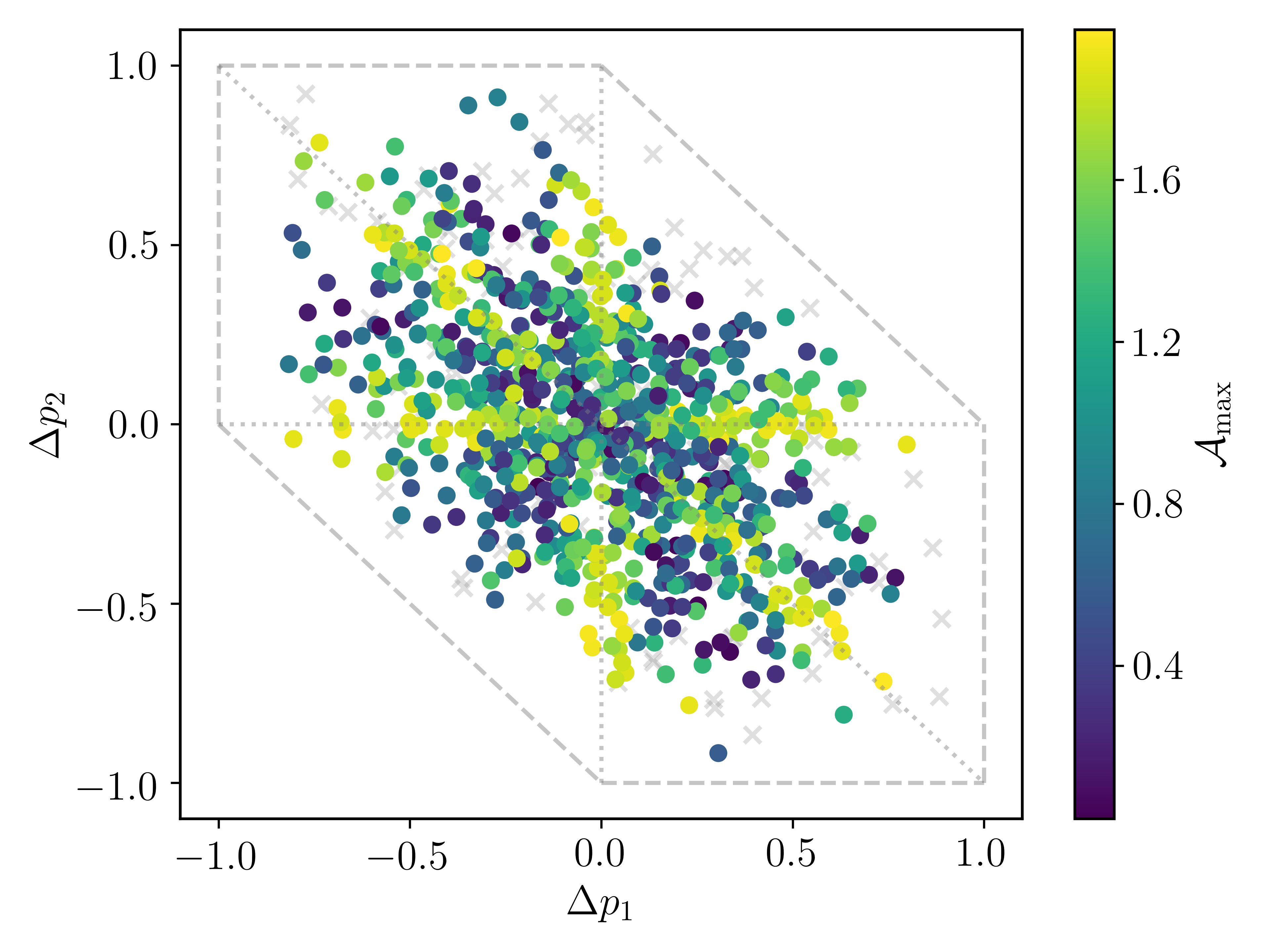}
	}
	\caption{Influence of the system-parameters on the maximal
		affinity $\mathcal A_{\rm{max}} \equiv \max_{t\in[0,T]} |\mathcal A(t)|$ of arbitrary sampled configurations for different speeds $T$. $\Delta p_i = p_i(0) - p_i(T)$ displays the change in the state densities. The colorbar represents $\mathcal A_{\rm{max}}$. The dashed line is defined by $\Delta p_i = \pm 1$ for one $i$. The dotted line represents $\Delta p_i = 0$ for a $i$. Grey crosses stand for configuration that failed to converge with our algorithm.}
	\label{fig:deltaPaffAll}
\end{figure}

	The faster the process is, the larger is the maximum
	affinity applied in the optimal process as shown in Fig. (\ref{fig:Amax}). While $\mathcal A_{\rm{max}} \equiv \max_{0\leq t \leq T}|\mathcal A(t)|$ remains below
	the value of $2$ as it should, there are combinations of
	initial and final densities for which the optimal affinity
	seems to reach this bound within about $2$ percent. In Fig. (\ref{fig:deltaPaffAll}), we show that the
	largest affinities are generated by those initial and final
	distributions that require to transport either the largest density, i.e., for which $\Delta p_i = p_i(0) - p_i(T) $ is approximately $\pm 1$
	for one pair of states, or for which $\Delta p_i \simeq 0$ for one state. 
	
	Comparing the configurations displayed in Fig. \ref{fig:deltaPaffAll}, we find that it becomes
	more difficult to obtain numerically convergent solutions the lower we set the allocated time $T$. 
	Particularly, we find configurations which tend to transport the largest densities, $\Delta p_i \to \pm 1$,
	to be numerically unstable (grey crosses). At present, it is unclear whether this is due to the numerical scheme \cite{SM}
	or whether there is a generic problem in the mathematical formulation, e.g., due to diverging derivatives of the driving functions or vanishing probabilities. Another question that remains open is whether there always exist a set of initial and final densities  that 
	saturate the bound of the affinity, Eq. (\ref{eq:sharpBound}), for a given time $T$. 
	From our numerical findings we expect that this is the case for configurations that transport the maximal densities, i.e., for $\Delta p_i \simeq \pm 1$. The larger the allocated time $T$, the closer the transported densities need to be to $\pm 1$ in order to saturate the bound.
	
	So far, with the parametrization (\ref{eq:Rates}), we have focused
	on a symmetric splitting of the driving over each forward and backward rate.
	In a more general setting, we now allow for a splitting that may be different 
	for each link.
	We can then parametrize the rates as
	\begin{equation}\begin{split}
	k_{ij}&= \kappa_{ij} \exp(\alpha_{ij} A_{ij})\\
	k_{ji} &= \kappa_{ji} \exp[(1-\alpha_{ij})A_{ji}]
	\end{split}
	\label{eq:Rates2}
	\end{equation}
	with $\kappa_{ij} = \kappa_{ji}$ and one structural parameter for each link given by $0 < \alpha_{ij}= \alpha_{ji} < 1$. Following the derivation in the symmetric case from above, it is straightforward to show that the bound (\ref{eq:bound}) becomes \cite{SM}
	\begin{equation}
	-\sum_{i=1}^{N_c} \frac{1}{\alpha_{i,i+1}} \leq \mathcal A_c \leq \sum_{i=1}^{N_c} \frac{1}{1-\alpha_{i,i+1}}~.
	\label{eq:symBound}
	\end{equation}
	If $\alpha_i = 1/2$ for all $i$, we reproduce Eq. (\ref{eq:bound}). 
	
	The more states a cycle has, the larger become our bounds. Naively extrapolating 
	to a cycle with infinitely many states, one might conclude that in such a continuum limit the affinity could diverge. Such an expectation would be in contrast with the established result that for a Langevin dynamics conservative driving, i.e., zero affinity achieves optimality \cite{aure11,aure12}. We finally show that our approach reproduces this continuum limit correctly. Let $\text dx$ denote a lattice spacing along a cycle. 
	We relabel the driving function of adjacent states in a cycle from $\varphi_{i,i+1}$ to $\varphi_{x,x+\text{d}x}$. For small lattice spacing, the affinity becomes
	\begin{equation}\begin{split}
			\mathcal A_c(t) &= \sum_{x\in\mathcal C} \varphi_{x,x+dx}(t) \approx \sum_{x\in\mathcal C} \varphi^\prime_{x,x}(t) \text dx  \approx \oint_{\mathcal C} \varphi^\prime_{x,x}(t) \text dx
	\end{split}\end{equation}
	where the prime denotes a derivative with respect to $x$. Here, we have used that $\varphi_{x,x}(t) = 0$ due to the antisymmetry of $\varphi_{ij}(t)$. Thus, the affinity approaches the contour integral over the spatial derivative of the driving function $\varphi^\prime_{x,x}(t)$ along the cycle. We can calculate the limiting value of this integral by dividing Eq. (\ref{eq:tanhSum}) by $-2$ and a Taylor expansion, according to
	\begin{equation}\begin{split}
			0&=\sum_{x\in\mathcal C} [\varphi_{x,x+\text dx}(t) + 2 \tanh \frac{\varphi_{x,x+\text dx}(t)}{2}]\\
			%&\approx \sum_x \bigg(\varphi_{x,x}+\varphi^\prime_{x,x}\text dx + 2\tanh \frac{\varphi_{x,x}}{2} + \frac{1}{\cosh^2\frac{\varphi_{x,x}}{2}} \varphi^\prime_{x,x} \text dx\bigg)\\
			&\approx 2 \sum_{x\in\mathcal C} \varphi^\prime_{x,x}(t)\text dx  \approx 2\oint_{\mathcal C} \varphi^\prime_{x,x}(t) \text dx .
		\end{split}\label{eq:zeroAff}\end{equation}
	Thus, the cycle affinity indeed has to vanish in the continuum limit. This finding also generalizes to the non-symmetrical splitting of the rates, Eq. (\ref{eq:Rates2}) \cite{SM}.
	
	In conclusion, we have proven that for discrete systems, optimal finite-time processes require non-conservative driving in marked contrast to the case of systems with continuous degrees of freedom. This result implies that
	driving a process, e.g., with unbalanced biochemical reactions can yield lower entropy production than by pumping the system through time-dependent
	modulations of energies and barriers. For each cycle in a multicyclic network, the maximum affinity remains bounded throughout the process, even if the allocated time approaches zero. For driving that affects forward and backward rates symmetrically, the bound depends only on the number of states of a cycle. For a non-symmetric splitting, a structural parameter enters the bound. Open theoretical problems include a proof of the tightness of the improved bound, Eq. (\ref{eq:sharpBound}), for all $T$ and a generalization of this improved bound to asymmetric splitting. For experiments,
	it remains a challenge to set up a system for
	which both types of driving, conservative and non-conservative one, can be implemented and
	quantitatively be compared with another at the 
	same time.

\textsl{Acknowledgments:}
	We thank Jann van der Meer and Timur Koyuk for stimulating discussions.

\onecolumngrid

\twocolumngrid

\bibliography{../../../Gitlab/Bibliography/refs,refs2} 

\begin{thebibliography}{31}%
\makeatletter
\providecommand \@ifxundefined [1]{%
 \@ifx{#1\undefined}
}%
\providecommand \@ifnum [1]{%
 \ifnum #1\expandafter \@firstoftwo
 \else \expandafter \@secondoftwo
 \fi
}%
\providecommand \@ifx [1]{%
 \ifx #1\expandafter \@firstoftwo
 \else \expandafter \@secondoftwo
 \fi
}%
\providecommand \natexlab [1]{#1}%
\providecommand \enquote  [1]{``#1''}%
\providecommand \bibnamefont  [1]{#1}%
\providecommand \bibfnamefont [1]{#1}%
\providecommand \citenamefont [1]{#1}%
\providecommand \href@noop [0]{\@secondoftwo}%
\providecommand \href [0]{\begingroup \@sanitize@url \@href}%
\providecommand \@href[1]{\@@startlink{#1}\@@href}%
\providecommand \@@href[1]{\endgroup#1\@@endlink}%
\providecommand \@sanitize@url [0]{\catcode `\\12\catcode `\$12\catcode
  `\&12\catcode `\#12\catcode `\^12\catcode `\_12\catcode `\%12\relax}%
\providecommand \@@startlink[1]{}%
\providecommand \@@endlink[0]{}%
\providecommand \url  [0]{\begingroup\@sanitize@url \@url }%
\providecommand \@url [1]{\endgroup\@href {#1}{\urlprefix }}%
\providecommand \urlprefix  [0]{URL }%
\providecommand \Eprint [0]{\href }%
\providecommand \doibase [0]{http://dx.doi.org/}%
\providecommand \selectlanguage [0]{\@gobble}%
\providecommand \bibinfo  [0]{\@secondoftwo}%
\providecommand \bibfield  [0]{\@secondoftwo}%
\providecommand \translation [1]{[#1]}%
\providecommand \BibitemOpen [0]{}%
\providecommand \bibitemStop [0]{}%
\providecommand \bibitemNoStop [0]{.\EOS\space}%
\providecommand \EOS [0]{\spacefactor3000\relax}%
\providecommand \BibitemShut  [1]{\csname bibitem#1\endcsname}%
\let\auto@bib@innerbib\@empty
%</preamble>
\bibitem [{\citenamefont {Andresen}(2011)}]{andr11}%
  \BibitemOpen
  \bibfield  {author} {\bibinfo {author} {\bibfnamefont {B.}~\bibnamefont
  {Andresen}},\ }\bibfield  {title} {\enquote {\bibinfo {title} {Current trends
  in finite-time thermodynamics},}\ }\href {\doibase 10.1002/anie.201001411}
  {\bibfield  {journal} {\bibinfo  {journal} {Angew. Chem. Int. Ed.}\ }\textbf
  {\bibinfo {volume} {50}},\ \bibinfo {pages} {2690--2704} (\bibinfo {year}
  {2011})}\BibitemShut {NoStop}%
\bibitem [{\citenamefont {Sekimoto}\ and\ \citenamefont
  {{S}asa}(1997)}]{seki97a}%
  \BibitemOpen
  \bibfield  {author} {\bibinfo {author} {\bibfnamefont {K.}~\bibnamefont
  {Sekimoto}}\ and\ \bibinfo {author} {\bibfnamefont {S.}~\bibnamefont
  {{S}asa}},\ }\bibfield  {title} {\enquote {\bibinfo {title} {Complementarity
  relation for irreversible process derived from stochastic energetics},}\
  }\href@noop {} {\bibfield  {journal} {\bibinfo  {journal} {J.\ Phys.\ Soc.\
  Jpn.}\ }\textbf {\bibinfo {volume} {66}},\ \bibinfo {pages} {3326} (\bibinfo
  {year} {1997})}\BibitemShut {NoStop}%
\bibitem [{\citenamefont {Schmiedl}\ and\ \citenamefont
  {Seifert}(2007)}]{schm07}%
  \BibitemOpen
  \bibfield  {author} {\bibinfo {author} {\bibfnamefont {T.}~\bibnamefont
  {Schmiedl}}\ and\ \bibinfo {author} {\bibfnamefont {U.}~\bibnamefont
  {Seifert}},\ }\bibfield  {title} {\enquote {\bibinfo {title} {Optimal
  finite-time processes in stochastic thermodynamics},}\ }\href {\doibase
  10.1103/PhysRevLett.98.108301} {\bibfield  {journal} {\bibinfo  {journal}
  {Phys.\ Rev.\ Lett.}\ }\textbf {\bibinfo {volume} {98}},\ \bibinfo {pages}
  {108301} (\bibinfo {year} {2007})}\BibitemShut {NoStop}%
\bibitem [{\citenamefont {Schmiedl}\ and\ \citenamefont
  {Seifert}(2008)}]{schm08}%
  \BibitemOpen
  \bibfield  {author} {\bibinfo {author} {\bibfnamefont {T.}~\bibnamefont
  {Schmiedl}}\ and\ \bibinfo {author} {\bibfnamefont {U.}~\bibnamefont
  {Seifert}},\ }\bibfield  {title} {\enquote {\bibinfo {title} {Efficiency at
  maximum power: An analytically solvable model for stochastic heat engines},}\
  }\href {\doibase 10.1209/0295-5075/81/20003} {\bibfield  {journal} {\bibinfo
  {journal} {EPL}\ }\textbf {\bibinfo {volume} {81}},\ \bibinfo {pages} {20003}
  (\bibinfo {year} {2008})}\BibitemShut {NoStop}%
\bibitem [{\citenamefont {Aurell}\ \emph {et~al.}(2012)\citenamefont {Aurell},
  \citenamefont {Mej\'{i}a-Monasterio},\ and\ \citenamefont
  {Muratore-Ginanneschi}}]{aure12}%
  \BibitemOpen
  \bibfield  {author} {\bibinfo {author} {\bibfnamefont {E.}~\bibnamefont
  {Aurell}}, \bibinfo {author} {\bibfnamefont {C.}~\bibnamefont
  {Mej\'{i}a-Monasterio}}, \ and\ \bibinfo {author} {\bibfnamefont
  {P.}~\bibnamefont {Muratore-Ginanneschi}},\ }\bibfield  {title} {\enquote
  {\bibinfo {title} {Boundary layers in stochastic thermodynamics},}\ }\href
  {\doibase 10.1103/PhysRevE.85.020103} {\bibfield  {journal} {\bibinfo
  {journal} {Phys.\ Rev.\ E}\ }\textbf {\bibinfo {volume} {85}},\ \bibinfo
  {pages} {020103} (\bibinfo {year} {2012})}\BibitemShut {NoStop}%
\bibitem [{\citenamefont {Esposito}\ \emph {et~al.}(2010)\citenamefont
  {Esposito}, \citenamefont {Kawai}, \citenamefont {Lindenberg},\ and\
  \citenamefont {van~den Broeck}}]{espo10a}%
  \BibitemOpen
  \bibfield  {author} {\bibinfo {author} {\bibfnamefont {M.}~\bibnamefont
  {Esposito}}, \bibinfo {author} {\bibfnamefont {R.}~\bibnamefont {Kawai}},
  \bibinfo {author} {\bibfnamefont {K.}~\bibnamefont {Lindenberg}}, \ and\
  \bibinfo {author} {\bibfnamefont {C.}~\bibnamefont {van~den Broeck}},\
  }\bibfield  {title} {\enquote {\bibinfo {title} {Finite time thermodynamics
  for a single level quantum dot},}\ }\href {\doibase
  10.1209/0295-5075/89/20003} {\bibfield  {journal} {\bibinfo  {journal} {EPL}\
  }\textbf {\bibinfo {volume} {89}},\ \bibinfo {pages} {20003} (\bibinfo {year}
  {2010})}\BibitemShut {NoStop}%
\bibitem [{\citenamefont {Diana}\ \emph {et~al.}(2013)\citenamefont {Diana},
  \citenamefont {Bagci},\ and\ \citenamefont {Esposito}}]{dian13}%
  \BibitemOpen
  \bibfield  {author} {\bibinfo {author} {\bibfnamefont {G.}~\bibnamefont
  {Diana}}, \bibinfo {author} {\bibfnamefont {G.~B.}\ \bibnamefont {Bagci}}, \
  and\ \bibinfo {author} {\bibfnamefont {M.}~\bibnamefont {Esposito}},\
  }\bibfield  {title} {\enquote {\bibinfo {title} {Finite-time erasing of
  information stored in fermionic bits},}\ }\href {\doibase
  10.1103/PhysRevE.87.012111} {\bibfield  {journal} {\bibinfo  {journal}
  {Phys.\ Rev.\ E}\ }\textbf {\bibinfo {volume} {87}},\ \bibinfo {pages}
  {012111} (\bibinfo {year} {2013})}\BibitemShut {NoStop}%
\bibitem [{\citenamefont {Zulkowski}\ and\ \citenamefont
  {DeWeese}(2014)}]{zulk11}%
  \BibitemOpen
  \bibfield  {author} {\bibinfo {author} {\bibfnamefont {Patrick~R.}\
  \bibnamefont {Zulkowski}}\ and\ \bibinfo {author} {\bibfnamefont
  {Michael~R.}\ \bibnamefont {DeWeese}},\ }\bibfield  {title} {\enquote
  {\bibinfo {title} {Optimal finite-time erasure of a classical bit},}\ }\href
  {\doibase 10.1103/PhysRevE.89.052140} {\bibfield  {journal} {\bibinfo
  {journal} {Phys. Rev. E}\ }\textbf {\bibinfo {volume} {89}},\ \bibinfo
  {pages} {052140} (\bibinfo {year} {2014})}\BibitemShut {NoStop}%
\bibitem [{\citenamefont {{C}rooks}\ and\ \citenamefont
  {{J}arzynski}(2007)}]{croo07}%
  \BibitemOpen
  \bibfield  {author} {\bibinfo {author} {\bibfnamefont {G.~E.}\ \bibnamefont
  {{C}rooks}}\ and\ \bibinfo {author} {\bibfnamefont {C.}~\bibnamefont
  {{J}arzynski}},\ }\bibfield  {title} {\enquote {\bibinfo {title} {Work
  distribution for the adiabatic compression of a dilute and interacting
  classical gas},}\ }\href {\doibase 10.1103/PhysRevE.75.021116} {\bibfield
  {journal} {\bibinfo  {journal} {Phys.\ Rev.\ E}\ }\textbf {\bibinfo {volume}
  {75}},\ \bibinfo {pages} {021116} (\bibinfo {year} {2007})}\BibitemShut
  {NoStop}%
\bibitem [{\citenamefont {Sivak}\ and\ \citenamefont {Crooks}(2012)}]{siva12}%
  \BibitemOpen
  \bibfield  {author} {\bibinfo {author} {\bibfnamefont {David~A.}\
  \bibnamefont {Sivak}}\ and\ \bibinfo {author} {\bibfnamefont {Gavin~E.}\
  \bibnamefont {Crooks}},\ }\bibfield  {title} {\enquote {\bibinfo {title}
  {Thermodynamic metrics and optimal paths},}\ }\href {\doibase
  10.1103/PhysRevLett.108.190602} {\bibfield  {journal} {\bibinfo  {journal}
  {Phys. Rev. Lett.}\ }\textbf {\bibinfo {volume} {108}},\ \bibinfo {pages}
  {190602} (\bibinfo {year} {2012})}\BibitemShut {NoStop}%
\bibitem [{\citenamefont {Zulkowski}\ \emph {et~al.}(2012)\citenamefont
  {Zulkowski}, \citenamefont {Sivak}, \citenamefont {Crooks},\ and\
  \citenamefont {DeWeese}}]{zulk12}%
  \BibitemOpen
  \bibfield  {author} {\bibinfo {author} {\bibfnamefont {Patrick~R.}\
  \bibnamefont {Zulkowski}}, \bibinfo {author} {\bibfnamefont {David~A.}\
  \bibnamefont {Sivak}}, \bibinfo {author} {\bibfnamefont {Gavin~E.}\
  \bibnamefont {Crooks}}, \ and\ \bibinfo {author} {\bibfnamefont {Michael~R.}\
  \bibnamefont {DeWeese}},\ }\bibfield  {title} {\enquote {\bibinfo {title}
  {Geometry of thermodynamic control},}\ }\href {\doibase
  10.1103/PhysRevE.86.041148} {\bibfield  {journal} {\bibinfo  {journal} {Phys.
  Rev. E}\ }\textbf {\bibinfo {volume} {86}},\ \bibinfo {pages} {041148}
  (\bibinfo {year} {2012})}\BibitemShut {NoStop}%
\bibitem [{\citenamefont {Muratore-Ginanneschi}(2013)}]{mura13}%
  \BibitemOpen
  \bibfield  {author} {\bibinfo {author} {\bibfnamefont {Paolo}\ \bibnamefont
  {Muratore-Ginanneschi}},\ }\bibfield  {title} {\enquote {\bibinfo {title} {On
  the use of stochastic differential geometry for non-equilibrium thermodynamic
  modeling and control},}\ }\href {\doibase 10.1088/1751-8113/46/27/275002}
  {\bibfield  {journal} {\bibinfo  {journal} {Journal of Physics A:
  Mathematical and Theoretical}\ }\textbf {\bibinfo {volume} {46}},\ \bibinfo
  {pages} {275002} (\bibinfo {year} {2013})}\BibitemShut {NoStop}%
\bibitem [{\citenamefont {Zulkowski}\ and\ \citenamefont
  {DeWeese}(2015)}]{zulk15}%
  \BibitemOpen
  \bibfield  {author} {\bibinfo {author} {\bibfnamefont {Patrick~R.}\
  \bibnamefont {Zulkowski}}\ and\ \bibinfo {author} {\bibfnamefont
  {Michael~R.}\ \bibnamefont {DeWeese}},\ }\bibfield  {title} {\enquote
  {\bibinfo {title} {Optimal control of overdamped systems},}\ }\href {\doibase
  10.1103/PhysRevE.92.032117} {\bibfield  {journal} {\bibinfo  {journal} {Phys.
  Rev. E}\ }\textbf {\bibinfo {volume} {92}},\ \bibinfo {pages} {032117}
  (\bibinfo {year} {2015})}\BibitemShut {NoStop}%
\bibitem [{\citenamefont {Sivak}\ and\ \citenamefont {Crooks}(2016)}]{siva16}%
  \BibitemOpen
  \bibfield  {author} {\bibinfo {author} {\bibfnamefont {David~A.}\
  \bibnamefont {Sivak}}\ and\ \bibinfo {author} {\bibfnamefont {Gavin~E.}\
  \bibnamefont {Crooks}},\ }\bibfield  {title} {\enquote {\bibinfo {title}
  {Thermodynamic geometry of minimum-dissipation driven barrier crossing},}\
  }\href {\doibase 10.1103/PhysRevE.94.052106} {\bibfield  {journal} {\bibinfo
  {journal} {Phys. Rev. E}\ }\textbf {\bibinfo {volume} {94}},\ \bibinfo
  {pages} {052106} (\bibinfo {year} {2016})}\BibitemShut {NoStop}%
\bibitem [{\citenamefont {B{\'e}rut}\ \emph {et~al.}(2012)\citenamefont
  {B{\'e}rut}, \citenamefont {Arakelyan}, \citenamefont {Petrosyan},
  \citenamefont {Ciliberto}, \citenamefont {Dillenschneider},\ and\
  \citenamefont {Lutz}}]{beru12}%
  \BibitemOpen
  \bibfield  {author} {\bibinfo {author} {\bibfnamefont {A.}~\bibnamefont
  {B{\'e}rut}}, \bibinfo {author} {\bibfnamefont {A.}~\bibnamefont
  {Arakelyan}}, \bibinfo {author} {\bibfnamefont {A.}~\bibnamefont
  {Petrosyan}}, \bibinfo {author} {\bibfnamefont {S.}~\bibnamefont
  {Ciliberto}}, \bibinfo {author} {\bibfnamefont {R.}~\bibnamefont
  {Dillenschneider}}, \ and\ \bibinfo {author} {\bibfnamefont {E.}~\bibnamefont
  {Lutz}},\ }\bibfield  {title} {\enquote {\bibinfo {title} {Experimental
  verification of {L}andauer's principle linking information and
  thermodynamics},}\ }\href {\doibase 10.1038/nature10872} {\bibfield
  {journal} {\bibinfo  {journal} {Nature}\ }\textbf {\bibinfo {volume} {483}},\
  \bibinfo {pages} {187--189} (\bibinfo {year} {2012})}\BibitemShut {NoStop}%
\bibitem [{\citenamefont {Jun}\ \emph {et~al.}(2014)\citenamefont {Jun},
  \citenamefont {Gavrilov},\ and\ \citenamefont {Bechhoefer}}]{jun14}%
  \BibitemOpen
  \bibfield  {author} {\bibinfo {author} {\bibfnamefont {Yonggun}\ \bibnamefont
  {Jun}}, \bibinfo {author} {\bibfnamefont {Mom\ifmmode
  \check{c}\else~\v{c}\fi{}ilo}\ \bibnamefont {Gavrilov}}, \ and\ \bibinfo
  {author} {\bibfnamefont {John}\ \bibnamefont {Bechhoefer}},\ }\bibfield
  {title} {\enquote {\bibinfo {title} {High-precision test of landauer's
  principle in a feedback trap},}\ }\href {\doibase
  10.1103/PhysRevLett.113.190601} {\bibfield  {journal} {\bibinfo  {journal}
  {Phys. Rev. Lett.}\ }\textbf {\bibinfo {volume} {113}},\ \bibinfo {pages}
  {190601} (\bibinfo {year} {2014})}\BibitemShut {NoStop}%
\bibitem [{\citenamefont {Proesmans}\ \emph {et~al.}(2020)\citenamefont
  {Proesmans}, \citenamefont {Ehrich},\ and\ \citenamefont
  {Bechhoefer}}]{proe20}%
  \BibitemOpen
  \bibfield  {author} {\bibinfo {author} {\bibfnamefont {Karel}\ \bibnamefont
  {Proesmans}}, \bibinfo {author} {\bibfnamefont {Jannik}\ \bibnamefont
  {Ehrich}}, \ and\ \bibinfo {author} {\bibfnamefont {John}\ \bibnamefont
  {Bechhoefer}},\ }\bibfield  {title} {\enquote {\bibinfo {title} {Finite-time
  landauer principle},}\ }\href {\doibase 10.1103/PhysRevLett.125.100602}
  {\bibfield  {journal} {\bibinfo  {journal} {Phys. Rev. Lett.}\ }\textbf
  {\bibinfo {volume} {125}},\ \bibinfo {pages} {100602} (\bibinfo {year}
  {2020})}\BibitemShut {NoStop}%
\bibitem [{\citenamefont {Woodside}\ and\ \citenamefont
  {Block}(2014)}]{wood14}%
  \BibitemOpen
  \bibfield  {author} {\bibinfo {author} {\bibfnamefont {Michael~T.}\
  \bibnamefont {Woodside}}\ and\ \bibinfo {author} {\bibfnamefont {Steven~M.}\
  \bibnamefont {Block}},\ }\bibfield  {title} {\enquote {\bibinfo {title}
  {{Reconstructing folding energy landscapes by single-molecule force
  spectroscopy}},}\ }\href {\doibase 10.1146/annurev-biophys-051013-022754}
  {\bibfield  {journal} {\bibinfo  {journal} {Annual Review of Biophysics}\
  }\textbf {\bibinfo {volume} {43}},\ \bibinfo {pages} {19--39} (\bibinfo
  {year} {2014})}\BibitemShut {NoStop}%
\bibitem [{\citenamefont {Camunas-Soler}\ \emph {et~al.}(2016)\citenamefont
  {Camunas-Soler}, \citenamefont {Ribezzi-Crivellari},\ and\ \citenamefont
  {Ritort}}]{camu16}%
  \BibitemOpen
  \bibfield  {author} {\bibinfo {author} {\bibfnamefont {Joan}\ \bibnamefont
  {Camunas-Soler}}, \bibinfo {author} {\bibfnamefont {Marco}\ \bibnamefont
  {Ribezzi-Crivellari}}, \ and\ \bibinfo {author} {\bibfnamefont {Felix}\
  \bibnamefont {Ritort}},\ }\bibfield  {title} {\enquote {\bibinfo {title}
  {{Elastic Properties of Nucleic Acids by Single-Molecule Force
  Spectroscopy}},}\ }\href {\doibase 10.1146/annurev-biophys-062215-011158}
  {\bibfield  {journal} {\bibinfo  {journal} {Annual Review of Biophysics}\
  }\textbf {\bibinfo {volume} {45}},\ \bibinfo {pages} {65--84} (\bibinfo
  {year} {2016})}\BibitemShut {NoStop}%
\bibitem [{\citenamefont {Ciliberto}(2017)}]{cili17}%
  \BibitemOpen
  \bibfield  {author} {\bibinfo {author} {\bibfnamefont {S.}~\bibnamefont
  {Ciliberto}},\ }\bibfield  {title} {\enquote {\bibinfo {title} {Experiments
  in stochastic thermodynamics: Short history and perspectives},}\ }\href
  {\doibase 10.1103/PhysRevX.7.021051} {\bibfield  {journal} {\bibinfo
  {journal} {Phys. Rev. X}\ }\textbf {\bibinfo {volume} {7}},\ \bibinfo {pages}
  {021051} (\bibinfo {year} {2017})}\BibitemShut {NoStop}%
\bibitem [{\citenamefont {Sinitsyn}\ and\ \citenamefont
  {Nemenman}(2007)}]{sini07}%
  \BibitemOpen
  \bibfield  {author} {\bibinfo {author} {\bibfnamefont {N.~A.}\ \bibnamefont
  {Sinitsyn}}\ and\ \bibinfo {author} {\bibfnamefont {I.}~\bibnamefont
  {Nemenman}},\ }\bibfield  {title} {\enquote {\bibinfo {title} {The berry
  phase and the pump flux in stochastic chemical kinetics},}\ }\href {\doibase
  10.1209/0295-5075/77/58001} {\bibfield  {journal} {\bibinfo  {journal} {EPL}\
  }\textbf {\bibinfo {volume} {77}},\ \bibinfo {pages} {58001} (\bibinfo {year}
  {2007})}\BibitemShut {NoStop}%
\bibitem [{\citenamefont {Rahav}\ \emph {et~al.}(2008)\citenamefont {Rahav},
  \citenamefont {Horowitz},\ and\ \citenamefont {{J}arzynski}}]{raha08}%
  \BibitemOpen
  \bibfield  {author} {\bibinfo {author} {\bibfnamefont {S.}~\bibnamefont
  {Rahav}}, \bibinfo {author} {\bibfnamefont {J.}~\bibnamefont {Horowitz}}, \
  and\ \bibinfo {author} {\bibfnamefont {C.}~\bibnamefont {{J}arzynski}},\
  }\bibfield  {title} {\enquote {\bibinfo {title} {Directed flow in
  nonadiabatic stochastic pumps},}\ }\href {\doibase
  10.1103/PhysRevLett.101.140602} {\bibfield  {journal} {\bibinfo  {journal}
  {Phys.\ Rev.\ Lett.}\ }\textbf {\bibinfo {volume} {101}},\ \bibinfo {pages}
  {140602} (\bibinfo {year} {2008})}\BibitemShut {NoStop}%
\bibitem [{\citenamefont {Chernyak}\ and\ \citenamefont
  {Sinitsyn}(2008)}]{cher08}%
  \BibitemOpen
  \bibfield  {author} {\bibinfo {author} {\bibfnamefont {V.~Y.}\ \bibnamefont
  {Chernyak}}\ and\ \bibinfo {author} {\bibfnamefont {N.~A.}\ \bibnamefont
  {Sinitsyn}},\ }\bibfield  {title} {\enquote {\bibinfo {title} {Pumping
  restriction theorem for stochastic networks},}\ }\href {\doibase
  10.1103/PhysRevLett.101.160601} {\bibfield  {journal} {\bibinfo  {journal}
  {Phys.\ Rev.\ Lett.}\ }\textbf {\bibinfo {volume} {101}},\ \bibinfo {pages}
  {160601} (\bibinfo {year} {2008})}\BibitemShut {NoStop}%
\bibitem [{\citenamefont {Fang}\ \emph {et~al.}(2019)\citenamefont {Fang},
  \citenamefont {Kruse}, \citenamefont {Lu},\ and\ \citenamefont
  {Wang}}]{fang19}%
  \BibitemOpen
  \bibfield  {author} {\bibinfo {author} {\bibfnamefont {Xiaona}\ \bibnamefont
  {Fang}}, \bibinfo {author} {\bibfnamefont {Karsten}\ \bibnamefont {Kruse}},
  \bibinfo {author} {\bibfnamefont {Ting}\ \bibnamefont {Lu}}, \ and\ \bibinfo
  {author} {\bibfnamefont {Jin}\ \bibnamefont {Wang}},\ }\bibfield  {title}
  {\enquote {\bibinfo {title} {Nonequilibrium physics in biology},}\ }\href
  {\doibase 10.1103/RevModPhys.91.045004} {\bibfield  {journal} {\bibinfo
  {journal} {Rev. Mod. Phys.}\ }\textbf {\bibinfo {volume} {91}},\ \bibinfo
  {pages} {045004} (\bibinfo {year} {2019})}\BibitemShut {NoStop}%
\bibitem [{\citenamefont {Mugnai}\ \emph {et~al.}(2020)\citenamefont {Mugnai},
  \citenamefont {Hyeon}, \citenamefont {Hinczewski},\ and\ \citenamefont
  {Thirumalai}}]{mugn20}%
  \BibitemOpen
  \bibfield  {author} {\bibinfo {author} {\bibfnamefont {Mauro~L.}\
  \bibnamefont {Mugnai}}, \bibinfo {author} {\bibfnamefont {Changbong}\
  \bibnamefont {Hyeon}}, \bibinfo {author} {\bibfnamefont {Michael}\
  \bibnamefont {Hinczewski}}, \ and\ \bibinfo {author} {\bibfnamefont
  {D.}~\bibnamefont {Thirumalai}},\ }\bibfield  {title} {\enquote {\bibinfo
  {title} {Theoretical perspectives on biological machines},}\ }\href {\doibase
  10.1103/RevModPhys.92.025001} {\bibfield  {journal} {\bibinfo  {journal}
  {Rev. Mod. Phys.}\ }\textbf {\bibinfo {volume} {92}},\ \bibinfo {pages}
  {025001} (\bibinfo {year} {2020})}\BibitemShut {NoStop}%
\bibitem [{\citenamefont {Aurell}\ \emph {et~al.}(2011)\citenamefont {Aurell},
  \citenamefont {Mej\'{i}a-Monasterio},\ and\ \citenamefont
  {Muratore-Ginanneschi}}]{aure11}%
  \BibitemOpen
  \bibfield  {author} {\bibinfo {author} {\bibfnamefont {E.}~\bibnamefont
  {Aurell}}, \bibinfo {author} {\bibfnamefont {C.}~\bibnamefont
  {Mej\'{i}a-Monasterio}}, \ and\ \bibinfo {author} {\bibfnamefont
  {P.}~\bibnamefont {Muratore-Ginanneschi}},\ }\bibfield  {title} {\enquote
  {\bibinfo {title} {Optimal protocols and optimal transport in stochastic
  thermodynamics},}\ }\href {\doibase 10.1103/PhysRevLett.106.250601}
  {\bibfield  {journal} {\bibinfo  {journal} {Phys.\ Rev.\ Lett.}\ }\textbf
  {\bibinfo {volume} {106}},\ \bibinfo {pages} {250601} (\bibinfo {year}
  {2011})}\BibitemShut {NoStop}%
\bibitem [{\citenamefont {Gawedzki}(2013)}]{gawe13}%
  \BibitemOpen
  \bibfield  {author} {\bibinfo {author} {\bibfnamefont {Krzysztof}\
  \bibnamefont {Gawedzki}},\ }\bibfield  {title} {\enquote {\bibinfo {title}
  {{Fluctuation Relations in Stochastic Thermodynamics}},}\ }\href
  {http://arxiv.org/abs/1308.1518} {\ ,\ \bibinfo {pages} {1--45} (\bibinfo
  {year} {2013})},\ \Eprint {http://arxiv.org/abs/1308.1518} {arXiv:1308.1518}
  \BibitemShut {NoStop}%
\bibitem [{\citenamefont {Dechant}\ and\ \citenamefont
  {Sakurai}(2019)}]{dech19}%
  \BibitemOpen
  \bibfield  {author} {\bibinfo {author} {\bibfnamefont {Andreas}\ \bibnamefont
  {Dechant}}\ and\ \bibinfo {author} {\bibfnamefont {Yohei}\ \bibnamefont
  {Sakurai}},\ }\bibfield  {title} {\enquote {\bibinfo {title} {{Thermodynamic
  interpretation of Wasserstein distance}},}\ }\href
  {http://arxiv.org/abs/1912.08405} {\ ,\ \bibinfo {pages} {1--8} (\bibinfo
  {year} {2019})},\ \Eprint {http://arxiv.org/abs/1912.08405}
  {arXiv:1912.08405} \BibitemShut {NoStop}%
\bibitem [{\citenamefont {Muratore-Ginanneschi}\ \emph
  {et~al.}({2013})\citenamefont {Muratore-Ginanneschi}, \citenamefont
  {Mej{\'i}a-Monasterio},\ and\ \citenamefont {Peliti}}]{mura12}%
  \BibitemOpen
  \bibfield  {author} {\bibinfo {author} {\bibfnamefont {P.}~\bibnamefont
  {Muratore-Ginanneschi}}, \bibinfo {author} {\bibfnamefont {C.}~\bibnamefont
  {Mej{\'i}a-Monasterio}}, \ and\ \bibinfo {author} {\bibfnamefont
  {L.}~\bibnamefont {Peliti}},\ }\bibfield  {title} {\enquote {\bibinfo {title}
  {Heat release by controlled continuous-time {M}arkov jump processes},}\
  }\href {\doibase {10.1007/s10955-012-0676-6}} {\bibfield  {journal} {\bibinfo
   {journal} {{Journal of Statistical Physics}}\ }\textbf {\bibinfo {volume}
  {{150}}},\ \bibinfo {pages} {{181--203}} (\bibinfo {year} {{2013}})},\
  \Eprint {http://arxiv.org/abs/1203.4062} {arXiv:1203.4062} \BibitemShut
  {NoStop}%
\bibitem [{\citenamefont {Seifert}(2012)}]{seif12}%
  \BibitemOpen
  \bibfield  {author} {\bibinfo {author} {\bibfnamefont {U.}~\bibnamefont
  {Seifert}},\ }\bibfield  {title} {\enquote {\bibinfo {title} {Stochastic
  thermodynamics, fluctuation theorems, and molecular machines},}\ }\href
  {\doibase 10.1088/0034-4885/75/12/126001} {\bibfield  {journal} {\bibinfo
  {journal} {Rep. Prog. Phys.}\ }\textbf {\bibinfo {volume} {75}},\ \bibinfo
  {pages} {126001} (\bibinfo {year} {2012})}\BibitemShut {NoStop}%
\bibitem [{SM()}]{SM}%
  \BibitemOpen
  \href@noop {} {}\bibinfo {note} {See Supplemental Material for a proof of the
  bounds, Eq. (\ref{eq:sharpBound}) and Eq. (\ref{eq:symBound}), and details on
  the numerical implementation.}\BibitemShut {Stop}%
\end{thebibliography}%


\begin{thebibliography}{2}%
\makeatletter
\providecommand \@ifxundefined [1]{%
 \@ifx{#1\undefined}
}%
\providecommand \@ifnum [1]{%
 \ifnum #1\expandafter \@firstoftwo
 \else \expandafter \@secondoftwo
 \fi
}%
\providecommand \@ifx [1]{%
 \ifx #1\expandafter \@firstoftwo
 \else \expandafter \@secondoftwo
 \fi
}%
\providecommand \natexlab [1]{#1}%
\providecommand \enquote  [1]{``#1''}%
\providecommand \bibnamefont  [1]{#1}%
\providecommand \bibfnamefont [1]{#1}%
\providecommand \citenamefont [1]{#1}%
\providecommand \href@noop [0]{\@secondoftwo}%
\providecommand \href [0]{\begingroup \@sanitize@url \@href}%
\providecommand \@href[1]{\@@startlink{#1}\@@href}%
\providecommand \@@href[1]{\endgroup#1\@@endlink}%
\providecommand \@sanitize@url [0]{\catcode `\\12\catcode `\$12\catcode
  `\&12\catcode `\#12\catcode `\^12\catcode `\_12\catcode `\%12\relax}%
\providecommand \@@startlink[1]{}%
\providecommand \@@endlink[0]{}%
\providecommand \url  [0]{\begingroup\@sanitize@url \@url }%
\providecommand \@url [1]{\endgroup\@href {#1}{\urlprefix }}%
\providecommand \urlprefix  [0]{URL }%
\providecommand \Eprint [0]{\href }%
\providecommand \doibase [0]{http://dx.doi.org/}%
\providecommand \selectlanguage [0]{\@gobble}%
\providecommand \bibinfo  [0]{\@secondoftwo}%
\providecommand \bibfield  [0]{\@secondoftwo}%
\providecommand \translation [1]{[#1]}%
\providecommand \BibitemOpen [0]{}%
\providecommand \bibitemStop [0]{}%
\providecommand \bibitemNoStop [0]{.\EOS\space}%
\providecommand \EOS [0]{\spacefactor3000\relax}%
\providecommand \BibitemShut  [1]{\csname bibitem#1\endcsname}%
\let\auto@bib@innerbib\@empty
%</preamble>
\bibitem [{\citenamefont {Hanke-Bourgeois}(2006)}]{hank06}%
  \BibitemOpen
  \bibfield  {author} {\bibinfo {author} {\bibfnamefont {M.}~\bibnamefont
  {Hanke-Bourgeois}},\ }\href {\doibase 10.1007/978-3-8351-9020-7} {\emph
  {\bibinfo {title} {Grundlagen der Numerischen Mathematik und des
  Wissenschaftlichen Rechnens}}}\ (\bibinfo {year} {2006})\BibitemShut
  {NoStop}%
\bibitem [{\citenamefont {Osada}\ \emph {et~al.}(2002)\citenamefont {Osada},
  \citenamefont {Funkhouser}, \citenamefont {Chazelle},\ and\ \citenamefont
  {Dobkin}}]{osad02}%
  \BibitemOpen
  \bibfield  {author} {\bibinfo {author} {\bibfnamefont {R.}~\bibnamefont
  {Osada}}, \bibinfo {author} {\bibfnamefont {T.}~\bibnamefont {Funkhouser}},
  \bibinfo {author} {\bibfnamefont {B.}~\bibnamefont {Chazelle}}, \ and\
  \bibinfo {author} {\bibfnamefont {D.}~\bibnamefont {Dobkin}},\ }\href@noop {}
  {\bibfield  {journal} {\bibinfo  {journal} {ACM Transactions on Graphics}\
  }\textbf {\bibinfo {volume} {21}},\ \bibinfo {pages} {807} (\bibinfo {year}
  {2002})}\BibitemShut {NoStop}%
\end{thebibliography}%

\end{document}

%\preprint{}
\title{Supplemental Material for "Optimality of non-conservative driving for finite-time processes with discrete states"}

\author{Benedikt Remlein and Udo Seifert}
\address{ II. Institut f\"ur Theoretische Physik, Universit\"at Stuttgart,
	70550 Stuttgart, Germany}
\date{\today}

\parskip 1mm

\nopagebreak

\maketitle

\section{Improved Bound for symmetric rates}
We proof Eq. (21) of the main text, i.e.,
\begin{equation}
|\mathcal A(t)| \leq 2(N-2),~~\forall t \in [0,T]
\end{equation}
where here $N\geq 3$ denotes the number of states of a cycle $\mathcal C$ in the network, $\mathcal A(t)$ the associated affinity and $T$ the total duration of the process. 
The key-idea of the proof is to ask for which affinities a solution can exist. For convenience we relabel the driving functions of adjacent states along a cycle as
\begin{equation}
x_i \equiv \varphi_{i,i-1}(t)
\end{equation}
for arbitrary but fixed $0 \leq t \leq T$.
Thus, Eq. (18) from the main text becomes
\begin{equation}
0 = \sum_{i=1}^{N} \left ( x_i + 2 \tanh \frac{x_i}{2}\right)
\end{equation}
and the affinity along the cycle reads
\begin{equation}
\mathcal A = \sum_{i=1}^{N}  x_i~.
\label{eq:Ax}\end{equation}
A possible solution to the variational problem has to satisfy both of these relations. We can thus write the first relation as
\begin{equation}
\mathcal A = - 2 \sum_{i=1}^{N}  \tanh \frac{x_i}{2} \equiv F_N(x_1,\ldots,x_N)~.
\label{eq:AFx}
\end{equation}
Let us now reparameterize the solution-space as
\begin{equation}\begin{split}
(x_1,\ldots,x_{k-1},x_k,x_{k+1},\ldots,x_N) \to (x_1,\ldots,x_{k-1},\mathcal A - \sum_{i\neq k} x_i,x_{k+1},\ldots,x_N)\end{split}
\end{equation}
using Eq. (\ref{eq:Ax}). This reflects that we are only interested in solutions that are on the $\mathcal A$-hyperplane for a given affinity. We now look for the maximum respectively minimum value of
\begin{equation}\begin{split}
f_{N-1}(x_1,\ldots,x_{k-1},x_{k+1},\ldots,x_N;\mathcal A) \equiv
F_N(x_1,\ldots,x_{k-1},\mathcal A - \sum_{i\neq k} x_i,x_{k+1},\ldots,x_N)
\end{split}
\end{equation}
which has to be at least $\mathcal A$ for a solution to exist. Be $L\gg1$ arbitrary and $x \equiv (x_1,\ldots,x_{k-1},x_{k+1},\ldots,x_N) \in (-L,L)^{N-1}$. 

\subsection{Local extrema}
At first, we consider the local maxima and minima. Setting the derivative of $f_{N-1}$ with respect to $x_i$ ($i\neq k$) to zero gives
\begin{equation}
\frac{1}{\cosh^2\frac{\mathcal A - \sum_{i\neq k} x_i^*}{2}} - \frac{1}{\cosh^2\frac{x_i^*}{2}} = 0~.
\end{equation}
Thus, we find for all local extrema $x^*$
\begin{equation}
\cosh \frac{x_i^*}{2} = \cosh \frac{\mathcal A - \sum_{i\neq k} x_i^*}{2}
\label{eq:Extrema}
\end{equation}
and the extreme-values are parameterized as
\begin{equation}\begin{split}
f_{N-1}^* &\equiv f_{N-1}(x^*;\mathcal A)\\& = - \frac{2}{\cosh\frac{\mathcal A - \sum_{i\neq k} x_i^*}{2}} \bigg[ \sum_{i\neq k} \sinh \frac{x_i^*}{2} + \sinh\frac{\mathcal A - \sum_{i\neq k} x_i^*}{2}\bigg]~.
\end{split}\end{equation}
This expression simplifies with $\sinh(x) = \sigma(x) \sqrt{\cosh^2(x) -1}$ and Eq. (\ref{eq:Extrema}) to
\begin{equation}\begin{split}
f_{N-1}^* &= - 2 ~\bigg|\tanh \frac{x_k^*}{2}\bigg| ~ \sum_i \sigma (x_i^*)\\
&= 2~\bigg|\tanh \frac{x_k^*}{2}\bigg| \{N_{-} -  N_{+}\}~.
\end{split}
\end{equation}
Here, $x_k^* \equiv \mathcal A - \sum_{i\neq k} x_i^*$, $N_\pm$ is the number of positive respectively negative $x_i^*$ and $\sigma$ denotes the sign-function. We now distinguish between positive and negative affinities.
\subsubsection{Positive Affinity}
Let us assume $\mathcal A > 0$. We want to examine when a solution to Eq. (\ref{eq:AFx}) can exist for positive affinity, hence we need to find an upper bound for $f_{N-1}^*$. The local extrema are elements of the $\mathcal A$-hyperplane, therefore we find
\begin{equation}
0 < \mathcal A = \sum_i x_i^*~.
\end{equation}
Thus, at least one of the \{$x_i^*$\} needs to positive, i.e. $N_+ \geq 1$. Since the number of states in the cycle satisfies $N = N_+ + N_-$, the number of negative $x_i^*$ is bounded by $N_- \leq N-1$. We can now bound the extreme value from above as
\begin{equation}\begin{split}
f_{N-1}^* &=2~\bigg|\tanh \frac{x_k^*}{2}\bigg| \{N_{-} -  N_{+}\}\\
& \leq 2 ~|\tanh \frac{x_k^*}{2}| \{N-1 - 1\}\\
& \leq 2 (N-2)
\end{split}\end{equation}
where we used $|\tanh(x)|\leq 1$ in the last line. Thus, the affinity must necessarily satisfy
\begin{equation}
\mathcal A\leq 2(N-2)
\end{equation}
for a possible solution inside $(-L,L)^{N-1}$ and $\mathcal A >0$.

\subsubsection{Negative Affinity}
Let us assume $\mathcal A < 0$.  We now need to find a lower bound on $f_{N-1}^*$ in order to make a statement about existence of a solution to Eq. (\ref{eq:AFx}). A negative affinity implies
\begin{equation}
\sum_i x_i^* < 0~.
\end{equation}
Thus, the number of negative $x_i^*$ is now bounded from below $N_- \geq 1$ and therefore $N_+ \leq N-1$. The minimum value is hence
\begin{equation}
f_{N-1}^* \geq -2(N-2)~.
\end{equation}
Therefore, the affinity needs to be bounded from below as
\begin{equation}
\mathcal A \geq- 2(N-2)
\end{equation}
by the same reasoning as in the previous section.\smallskip

In total, we find that the affinity needs to satisfy
\begin{equation}
|\mathcal A| \leq 2(N-2)
\label{eq:boundStrong}
\end{equation}
to obtain a solution inside $(-L,L)^{N-1}$. To make the proof complete, we examine the behavior on the boundary $\partial (-L,L)^{N-1}$.

\subsection{Extrema On The Boundary}
We claim that Eq. (\ref{eq:boundStrong}) also holds on the boundary and prove it by induction.
\subsubsection{Initial Step: $N = 3$}
\begin{figure}
	\includegraphics[width=0.75\textwidth]{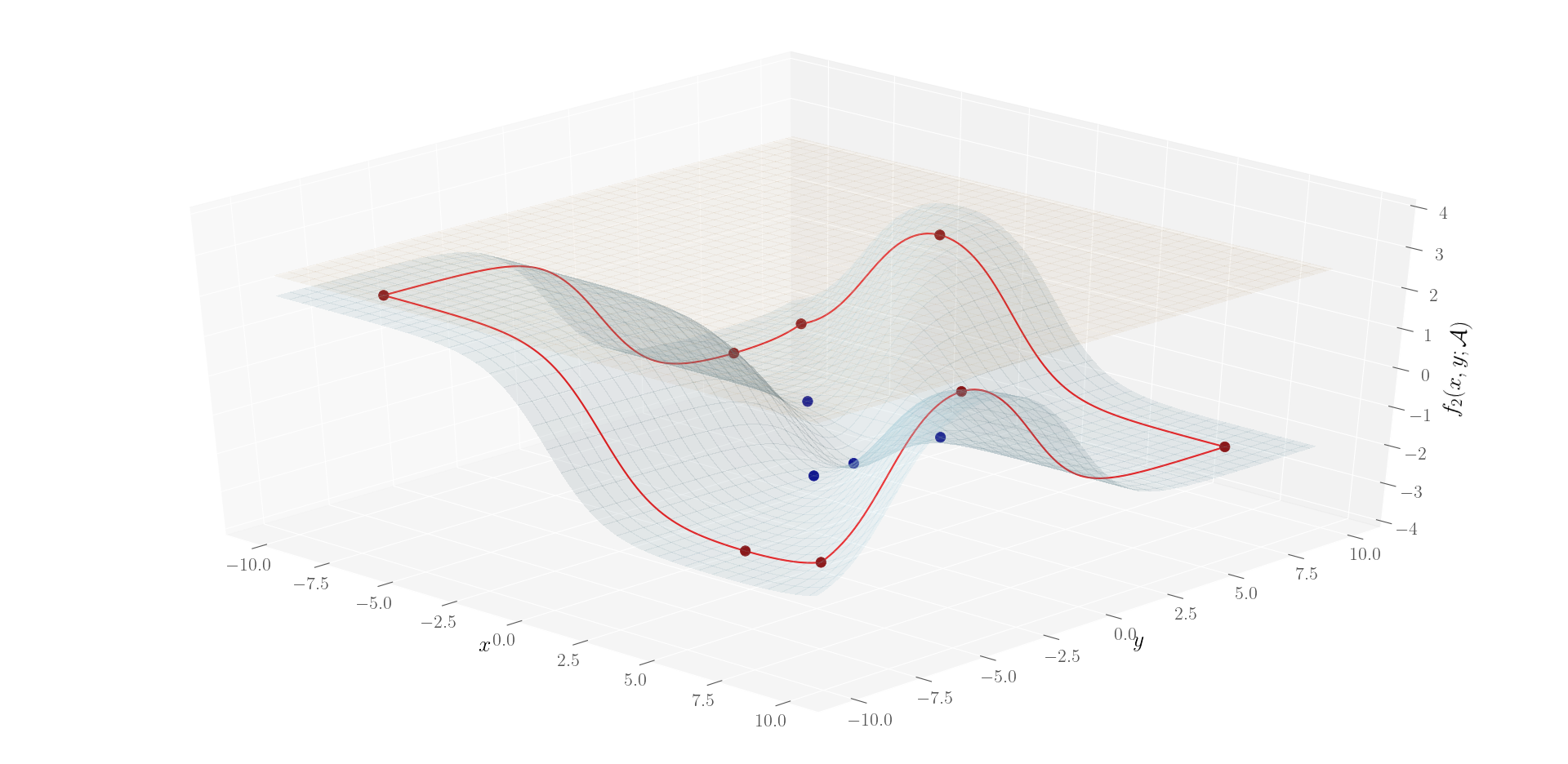}
	\caption{Illustration of the explicit calculations for the 3 state system. The blue graph is $f_2(x_1,x_2;\mathcal A)$ and the red surface is $\mathcal A = x_1+x_2+x_3$. The thick red line represents $f_2$ along the boundary $\partial (-L,L)^2$. The red dots display the 8 boundary-points discussed in the text. Blue dots represent the local extrema inside $(-L,L)^2$. In this example we set $\mathcal A = 2.5$ and $L = 8$, thus, there's no intersection between the $\mathcal A$-plane and the graph of $f_2$.}
	\label{fig:proofSketch}
\end{figure}

We calculate the bounds explicitly for $N = 3$. $x_3$ is parameterized as $x_3 = \mathcal A -x_1 - x_2$ and $\partial (-L,L)^2$ as 4 lines and 4 points, see Fig. \ref{fig:proofSketch}.
\begin{enumerate}
	\item $x_1 = -L$ \& $x_2 \in (-L,L)$:\\ We find a local extremum at 
	\begin{equation}
	x_2^- = \frac{\mathcal A +L}{2}
	\end{equation}
	with value
	\begin{equation}
	f_2(-L,x_2^-) = -2 \bigg  [ 2 \tanh \frac{L+\mathcal A }{2}-\tanh \frac L2 \bigg]\geq-2
	\end{equation}
	\item $x_1 = +L$ \& $x_2 \in (-L,L)$:\\ We find a local extremum at 
	\begin{equation}
	x_2^+ = \frac{\mathcal A -L}{2}
	\end{equation}
	with value
	\begin{equation}
	f_2(L,x_2^+) = 2 \bigg  [  2 \tanh \frac{L-\mathcal A}{2}-\tanh \frac L2 \bigg]\leq2
	\end{equation}
	
	\item $x_1 \in (-L,L)$ \& $x_2 = -L$:\\ We find a local extremum at 
	\begin{equation}
	x_1^- = \frac{\mathcal A + L}{2}
	\end{equation}
	with value
	\begin{equation}
	f_2(x_1^-,-L) = -2 \bigg  [  2 \tanh \frac{L+\mathcal A}{2}-\tanh \frac L2 \bigg]\geq-2
	\end{equation}
	
	\item $x_1 \in (-L,L)$ \& $x_2 = +L$:\\ We find a local extremum at 
	\begin{equation}
	x_1^+ = \frac{\mathcal A - L}{2}
	\end{equation}
	with value
	\begin{equation}
	f_2(x_1^+,L) = 2 \bigg  [  2 \tanh \frac{L-\mathcal A}{2}-\tanh \frac L2 \bigg]\leq2
	\end{equation}
	
	\item $x_1 = L$ \& $x_2 = L$: 
	\begin{equation}
	f_2(L,L) = -2 \bigg  [ 2\tanh \frac L2 - \tanh \frac{2L-\mathcal A}{2}\bigg]\geq -2 
	\end{equation}
	\item $x_1 = -L$ \& $x_2 = -L$: 
	\begin{equation} 
	f_2(-L,-L) = 2 \bigg  [ 2\tanh \frac L2 - \tanh \frac{2L-\mathcal A}{2}\bigg]\leq2 
	\end{equation}
	\item $x_1 = -L$ \& $x_2 = L$:
	\begin{equation}\begin{split}
	f_2(-L,L) &= -2 \tanh \frac{\mathcal A}{2} \\
	|f_2(-L,L)| &\leq 2
	\end{split}
	\end{equation}
	\item $x_1 = L$ \& $x_2 = -L$: 
	\begin{equation}\begin{split}
	f_2(L,-L) &= -2 \tanh \frac{\mathcal A}{2} \\
	|f_2(L,-L)| &\leq 2
	\end{split}
	\end{equation}
	
\end{enumerate}

Thus, we find on $\partial (-L,L)^2$ that 
\begin{equation}
|f_2(x_1,x_2)| \leq 2 = 2~(N -2)
\end{equation}
for $N=3$.

\subsubsection{Inductive Step}
Let us assume that for an arbitrary $N \geq 3$ 
\begin{equation}
|F_{N}(x_1,\ldots,x_N)| \leq 2(N-2)~.
\end{equation}
holds true. Let $k \in \{1,2,\ldots, N+1\}$ be arbitrary and set $x_k = \pm L$. Let us now write
\begin{equation}\begin{split}
F_{N+1}(x_1,\ldots, x_{k-1},\pm L,x_{k+1},\ldots,x_{N+1}) = - 2 \sum_{i\neq k} \tanh \frac{x_i}{2} \mp 2 \tanh \frac L2 \\ = F_N(x_1,\ldots, x_{k-1},x_{k+1},\ldots,x_{N+1})\mp 2 \tanh \frac L2~.\end{split}
\end{equation}
Thus, we find by assumptions and using $|\tanh x |\leq 1$
\begin{equation}\begin{split}  
\bigg |F_{N+1}(x_1,\ldots, x_{k-1},\pm L,x_{k+1},\ldots,x_{N+1})\bigg | &\leq \bigg |F_N(x_1,\ldots, x_{k-1},x_{k+1},\ldots,x_{N+1})\bigg | + 2 \bigg |\tanh \frac L2 \bigg | \\ & \quad \leq 2(N-2) + 2 = 2(N-1)~.\end{split}
\end{equation}
This completes the proof by induction since $k$ is arbitrary,

\subsection{Concluding Remark}
We specified the length $L$ of the $N-1$-cube as arbitrary large. In fact, we need to be more precise for all statements to be true. In order to include all local extrema, we need $L > \max_i |x_i^*|$. We suppressed this conditions since we are interested in the limit $L\to \infty$ and by definition, setting the derivative zero will find only finite $x_i^*$. Moreover, all arguments in the explicit calculations for $N = 3$ are only true for $L > |\mathcal A|$ which again can be neglected since we already know that $|\mathcal A| \leq 2N < \infty$ and we examine $L\to \infty$. The important fact is that all derived bounds are independent on $L\gg1$.

\section{Numerical scheme}
In this section, we describe the numerical scheme that we implemented in Python to solve the Euler-Lagrange equations, Eq. (7,16,17) in the main text.
The imposed boundary value problem is of the following form:
\begin{equation}\begin{split}
\partial_t \phi_i(t) &= f_i(\{\phi_i(t)\},\{\eta_i(t)\},\{\varphi_{ij}(t)\})\\
\partial_t \eta_i(t) &= g_i(\{\phi_i(t)\},\{\eta_i(t)\},\{\varphi_{ij}(t)\})\\
0&= r_{ij}(\{\phi_i(t)\},\{\eta_i(t)\},\{\varphi_{ij}(t)\})
\label{eq:AODE}
\end{split}\end{equation}
with given boundary values $\{\phi_i(0) = \sqrt{p_i^0}\}$ and $\{\phi_i(T) = \sqrt{p_i^T}\}$. The relation above is a system of ordinary differential equations coupled with a non-linear, algebraic constrain. For the concrete set-up, the algebraic equation is given through the variation with respect to $\{\varphi_{ij}(t)\}$
\begin{equation}\begin{split}
r_{ij}(t) = \eta_i(t) \phi_j(t)  - \eta_j(t)\phi_i(t)+  4 \phi_i(t)\phi_j(t) \bigg[ \tanh \frac{\varphi_{ij}(t)}{2} + \frac{\varphi_{ij}(t)}{2}\bigg]~.
\end{split}\label{eq:varphi}\end{equation}
We solve Eq. (\ref{eq:AODE}) using a shooting method \cite[chapter 87]{hank06} where we treat the $\{\eta_i(0)\}$ as free shooting-parameter. The numerical scheme can be summarized as follows:
\begin{enumerate}
	\item Make an initial guess for $\{\eta_i(0)\}$
	\item Solve Eq. (\ref{eq:varphi}) for $\{\varphi_{ij}(0)\}$ with given $\{\phi_i(0)\}$ and $\{\eta_i(0)\}$.
	\item Calculate \[ \begin{pmatrix}\{\phi_i(t+\Delta t)\}\\\{\eta_i(t+\Delta t)\}\end{pmatrix} \approx \text{G-2}(\{\phi_i(t)\},\{\eta_i(t)\},\{\varphi_{ij}(t)\})\]
	\item Update $\{\varphi_{ij}(t+\Delta t)\}$ by solving $0 = r_{ij}(\{\phi_i(t+\Delta t)\},\{\eta_i(t+\Delta t)\},\{\varphi_{ij}(t+\Delta t)\})$.
	\item Repeat step $3$ and $4$ until $T$ is reached.
	\item Compare the numerically obtained values for $\{\phi_i(T)\}$ with the given ones $\{\phi_i^T\equiv \sqrt{p_i^T}\}$:
	\begin{itemize}
		\item If $\phi_i(T) \approx \phi_i^T$ within a tolerance of $\Delta t$, accept the solution.
		\item If $|\phi_i(T) - \phi_i^T | > \Delta t$ for at least one $i$, start from $1$ with another initial guess.
	\end{itemize}
\end{enumerate}
We used the implicit 2-step Gauß-scheme ($\text{G-2}$) \cite[chapter 78 ]{hank06} to integrate the ODE-part of Eq. (\ref{eq:AODE}) and the SciPy-root function to solve Eq. (\ref{eq:varphi}). For updating the initial guess for $\{\eta_i(0)\}$ we also used the root function of SciPy. We iterated this scheme up to 20 times to find convergent solutions. The step size was chosen as $\Delta t = T/n$ where we kept $n = 10000$ constant.

\section{Randomly Generated Configurations}
\begin{figure}
	\includegraphics[width=0.55\textwidth]{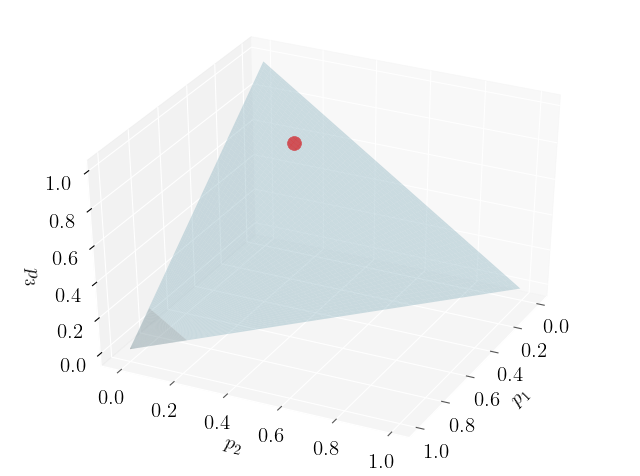}
	\caption{Random point (red dot) in the allowed region of the probabilities-phase space, which is defined by the the constrains $0<p_i<1$ and $\sum_i p_i = 1$ (blue area).}
	\label{fig:randomSketch}
\end{figure}

We now describe how we generated the random set of initial and final distributions. In order to obtain uniformly distributed densities from the allowed parameter-ranges of the 3 states system, we used the following parametrization \cite[Eq. (1)]{osad02}
\begin{equation}\begin{split}
\begin{pmatrix} p_1\\p_2\\p_3\end{pmatrix}(u,v) &= (1-\sqrt u) ~\vec e_1 + \sqrt u (1-v)~ \vec e_2 + v \sqrt u ~\vec e_3
\end{split}\end{equation}
with $\vec e_i$ the unit-vector in the $i$-th direction, see Fig. \ref{fig:randomSketch}. Choosing $u,v$ uniformly distributed between $0$ and $1$ results in the vector $\vec p = (p_1,p_2,p_3)^T$ to be uniformly distributed over the surface area of the triangle of allowed state densities.\smallskip

We generated 999 configurations. We could not find a numerically stable solution for 13 configuration for an allocated time $T = 1$, not for 61 for $T=0.1$ and not for 112 for $T = 0.01$.

\section{Affinity bound for non-symmetrical splitting}
We determine the equivalent of the bound, Eq. (20), on the affinity for the case of a non-symmetrical splitting of the jump-rates, i.e., for
	\begin{equation}\begin{split}
k_{ij}(t)&= \kappa_{ij} \exp(\alpha_{ij} A_{ij}(t))\\
k_{ji}(t) &= \kappa_{ji} \exp[(1-\alpha_{ij})A_{ji}(t)]
\end{split}
\label{eq:Rates2}
\end{equation}
with $\kappa_{ij} = \kappa_{ji} = const.$ and $0<\alpha_{ij} = \alpha_{ji} < 1$.

The Lagrange-function is given by
\begin{equation}
\begin{split}
L(t) &=\sum_{i\neq j} p_i(t) \kappa_{ij} \exp[\alpha_{ij} A_{ij}(t)] \ln\{\frac{p_i(t)}{p_j(t)} \exp[A_{ij}(t)]\}\\
& \quad +  \sum_i \eta_i(t)\{\partial_t p_i(t) - \sum_{j\neq i}[\kappa_{ji} \exp((1-\alpha_{ij})A_{ji}(t))p_j(t) -\kappa_{ij} \exp(\alpha_{ij} A_{ij}(t))p_i(t)]\}~.
\end{split}
\end{equation}
To find an expression for a cycle affinity, we need to evaluate the variation with respect to $A_{ij}(t)$. From $\delta L/\delta A_{ij}(t) = 0$ ($i<j$) we find
\begin{equation}
A_{ij}(t) = - \ln \frac{p_i(t)}{p_j(t)} - \eta_i(t) + \eta_j(t) - \frac{p_i(t) \exp[A_{ij}(t)]-p_j(t)}{\alpha_{ij}\exp[A_{ij}(t)] p_i(t) + (1-\alpha_{ij})p_j(t)}~.
\end{equation}
We sum this relation over an arbitrary cycle $\mathcal C$ in the network, to find for the corresponding affinity
\begin{equation}
\mathcal A_c(t) \equiv \sum_{(i,j)\in\mathcal C} A_{ij}(t)= - \sum_{(i,j)\in\mathcal C}\frac{p_i(t) \exp[A_{ij}(t)]-p_j(t)}{\alpha_{ij}\exp[A_{ij}(t)] p_i(t) + (1-\alpha_{ij})p_j(t)}~.
\label{eq:affinity}
\end{equation}
This relation simplifies to
\begin{equation}
\mathcal A_c(t) = - \sum_{(i,j)\in\mathcal C}\frac{1}{\alpha_{ij} + (1-\alpha_{ij})\exp[-A_{ij}(t)]p_j(t)/p_i(t)}
+ \sum_{(i,j)\in\mathcal C}\frac{1}{\alpha_{ij}\exp[A_{ij}(t)] p_i(t)/p_j(t) + (1-\alpha_{ij})}~.
\label{eq:simpleAffinity}
\end{equation}
Since all terms in the denominator are positive, we use $ 1/(a+b) \leq 1/a$ for $a,b >0$ to derive the bound, Eq. (23), on the cycle affinity
\begin{equation}
-\sum_{(i,j)\in\mathcal C}\frac{1}{\alpha_{ij}} \leq \mathcal A_c(t) \leq \sum_{(i,j)\in\mathcal C}\frac{1}{1-\alpha_{ij}}~.
\label{eq:alphaBound}
\end{equation}

Alternatively, we can rewrite Eq. (\ref{eq:simpleAffinity}) as
\begin{equation}
\mathcal A_c(t) = -2\sum_{(i,j)\in\mathcal C} \frac{\tanh\frac{\varphi_{ij}(t)}{2}}{1 - (1-2\alpha_{ij}) \tanh\frac{\varphi_{ij}(t)}{2}}
\end{equation}
with $\varphi_{ij}(t) = A_{ij}(t) + \ln p_i(t) - \ln p_j(t)$, which is a direct generalization of the symmetric case, Eq. (19), described in the main text. Eq. (\ref{eq:alphaBound}) follows from the monotonic behavior in $\varphi_{ij}(t)$ of the summands for given $\alpha_{ij}$.  

\section{Continuum Limit for non-symmetrical splitting}
We derive the continuum limit of a cycle affinity for rates parameterized as Eq. (\ref{eq:Rates2}). Relabeling adjacent states in a cycle from ($i$,$j$) to ($x$,$x+\text dx$) with a lattice spacing $\text dx$ leads to
\begin{equation}
\mathcal A_c(t) = \sum_{x\in\mathcal C} A_{x,x+ \text dx}(t)~.
\end{equation}
In the limit $\text dx \to 0$ we find by a Taylor-expansion
\begin{equation}
\mathcal A_c(t) \approx \sum_{x\in\mathcal C} A^\prime_{x,x}(t)\text dx \approx \oint_{\mathcal C} A^\prime_{x,x}(t) \text dx
\label{eq:affCont}
\end{equation}
where the prime denotes a spatial derivative.
The limit value of this contour integral is dictated by a Taylor-expansion of Eq. (\ref{eq:affinity}) for small $\text dx$
\begin{equation}
\mathcal A_c(t) \approx \sum_{x\in\mathcal C} \frac{p_x^\prime(t)}{p_x(t)}\text dx - \sum_{x\in\mathcal C} A^\prime_{x,x}(t)\text dx \approx \oint_{\mathcal C} \text d\log p_x(t) - \oint_{\mathcal C} A^\prime_{x,x}(t) \text dx~.
\end{equation}
We now combine this relation with Eq. \ref{eq:affCont} to obtain
\begin{equation}
\mathcal A_c(t)  \approx  \frac 12\oint_{\mathcal C} \text d\log p_x(t) = 0~.
\end{equation}
In the last step, we have used that the closed contour integral over a total differential vanishes.

\bibliography{../../refs,refs2}